\documentclass[twocolumn,10pt,a4paper]{article}
\usepackage{amsmath,amsfonts,mathtools,amssymb}
\usepackage{graphicx}
\usepackage{varioref}
\usepackage{cuted}  % for \begin{strip} ... \end{strip}
\usepackage[round]{natbib}
\usepackage{titlesec}
\usepackage{caption}

\titleformat{\section}
  {\normalfont\centering} % format
  {\bfseries\thesection} % label
  {0.7em} % sep
  {\bfseries} % before code 
  [] % after code

\titleformat{\subsection}
  {\normalfont\centering} % format
  {\itshape\thesubsection} % label
  {0.7em} % sep
  {\itshape} % before code 
  [] % after code

\newcommand\ds\displaystyle

\newcommand\dd{\,\mathrm{d}}
\DeclareMathOperator\dirac{\delta}
\DeclareMathOperator\Eint{E_1}
\newcommand\Esp{\mathbb{E}}
\newcommand\Exp{\mathcal{E}}
\newcommand\e{\mathrm{e}}
\newcommand\ue{\hat{\mathbf e}}
\newcommand\gev{g\sb{\rm e}}
\newcommand\go{g\sb{\rm o}}
\newcommand\ii{\mathrm{i}}
\DeclareMathOperator\besselI{\mathrm{I}}
\newcommand\Idist{\mathcal{I}}
\newcommand\lp{\ell\sp\star}
\let\Re\relax
\DeclareMathOperator\Re{Re}
\newcommand\vr{\mathbf{r}}
\renewcommand\Ref{\mathcal{R}} % Necessary for varioref
\newcommand\dir{u}
\newcommand\uu{\hat{\mathbf{u}}}
\newcommand\vv{\mathbf v}
\DeclareMathOperator\Var{Var}
\newcommand\Xg{\mathsf{X}}
\newcommand\XL{\mathcal{X}}
\newcommand\TL[1]{\widehat{#1}}
\newcommand\CTPRW{continous-time persistent random walk}
\newcommand\RTE{radiative transfer equation}

\def\titrecourt{Isotropic radiative transfer as a phase space process}
\def\titre{\titrecourt:\\
Lorentz covariant Green's functions and first-passage times}

\begin{document}

\title{\titre}
\author{Vincent Rossetto}
%\email{vincent.rossetto@grenoble.cnrs.fr}
%\institute{Univ. Grenoble Alpes, CNRS, LPMMC, 38000 Grenoble, France}
\date{\normalsize Univ. Grenoble Alpes, CNRS, LPMMC, 38000 Grenoble, France}
\maketitle

\vskip-4cm

%\abstract{
%\begin{abstract}
\begin{strip}
\begin{quote}
The solutions of the \RTE{},
known for the energy density, do not satisfy the fundamental
transitivity property for Green's functions expressed by 
Chapman-Kolmogorov's relation. I~show that this property is
retrieved by considering the radiance distribution in phase space.
Exact solutions are obtained in one and two dimensions
as probability density functions
of \CTPRW{s}, which Fokker-Planck equation is the \RTE{}.
The expected property of Lorentz covariance is verified.
I~also discuss the measured signal from a pulse source in
one dimension, which is a first-passage time
distribution, and unveil an effective random delay when the
pulse is emitted away from the observer.
%}
%\end{abstract}
\end{quote}
\end{strip}

%\begin{document}

\section{Introduction}
The physics of waves propagating in a cloud of scatterers,
generically called \emph{radiative transfer},
is a topic of intense research in 
astrophysics \citep{ishimaru}, meteorology \citep{marshakdavis}, 
medical imaging \citep{asllanaj2014},
seismology \citep{shang1988} and computer graphics \citep{jakob2010}.
While specific questions are addressed in these domains of application,
some fundamental aspects apply to all of them such as the spreading
of energy into space, which is governed by Chandrasekhar's \RTE{} 
\citeyearpar{chandrasekhar}. 
Chandrasekhar's equation describes the spatial and temporal evolution
of the \emph{radiance} (or the \emph{luminance})
$\phi$ as a function of the position~$\vr$, the direction of propagation~$\uu$
and time.
The radiance (luminance) is the (luminous) energy flux density. 
In a non-absorbing medium, the \RTE{} is
\begin{equation}
 \partial_t\phi
  +c\uu\cdot\nabla\phi
  +\frac{c}\ell \phi=\frac c\ell\Idist[\phi]+S(t,\,\vr,\,\uu), 
\label{RTE}
\end{equation}
where $c$ is the speed of light, $\ell$ is the mean free path,
and $S$ is a source term.
The integral operator $\Idist$ is the emission term which 
in the case of isotropic scattering is the average of $\phi$
with respect to~$\uu$. Only isotropic scattering is considered in this Letter.
I will also neglect absorption and coherent effects such as interferences.

The solutions of the \RTE{}~\eqref{RTE} 
strongly depend on the spatial dimension.
In one dimension, the \RTE{} reduces to 
the telegrapher's equation, a second order linear
partial differential equation
arising from electric transport theory, dating back to 
\citet{thomson1855}. The telegrapher's equation 
has a wide range of applications~\citep{weiss2002}.
It was solved %a century later
by \citet{hemmer1961} as he studied a modified
version of Smoluchowski's diffusion equation \citep{brinkman1956,sack1956}.
%Morse and Feschbach studied this equation in the
%context of heat transfer \cite[p. 865]{morsefeschbach1}.
The solution of the \RTE{} in two dimensions 
was obtained by seismologists
\citet{shang1988} and \citet{sato1993}. In three dimensions no exact
solutions have been derived despite the important efforts put to the task.
Several approximations have nonetheless been obtained, the most notable
by \citet{paasschens1997}.

As in all transport phenomena, 
it seems natural that the elementary solutions to the \RTE{} have
the property of \emph{transitivity}, which means that the spatial
distribution of energy at a time $t_2$ can be deduced from the
energy distribution at a previous time $t_1<t_2$ using the distribution of
energy from a pulse after a time $t_2-t_1$, but this is not the case
\citep{dunkel2009}; in other words, as it will be shown 
in the Section \ref{sec:cpr} of this Letter, 
the elementary solutions from 
Hemmer, and Shang and Gao are not Green's functions.

In this Letter, I show that expressing the solutions in
phase space rather than in position space are the Green's functions
in one and two dimensions. I show that these solutions
are Lorentz covariant and have the transitivity property.
The following Section 
introduces \CTPRW{s} (CTPRW) and their relation to the \RTE{}.
In the Section \ref{sec:sol1}, 
I compute the mean free path in one dimension, 
in a frame in relative motion with the cloud of scatterers, 
and I show that it depends on the direction of propagation.
This asymmetry suggests a reinterpretation of the phase space solution 
for asymmetric scattering  
as the solution for the symmetric case transformed by a Lorentz boost.
In the Section \ref{sec:cpr}, I show that
the CTPRW as a phase space process has the Markov property and 
discuss why the solution expressed in position space does not
satisfy Chapman-Kolmogorov's relation.
The following Section is dedicated to the first-passage time properties 
of the CTPRW: I show that an effective new "flip" process arises, 
of which I provide elementary properties. 
In the Section \ref{sec:lc2}, I show that the
already known phase space solution in two dimensions is Lorentz
covariant, a fact that was not hitherto known.
Lastly, before concluding, I discuss the Brownian limit $c\to\infty$ 
of the investigated random walks and show that the 
phase space Green's functions asymptotically approach a
Gaussian in this limit, which only depends on position.

\section{Persistent random walks}
In 1951, Goldstein remarked that the telegrapher's equation
is also the Fokker-Planck equation of a persistent random walk
\citep{goldstein1951}. Therefore, the probability density function 
of a persistent random walk obeys the telegrapher's equation, in the
same manner that the probability density function of the Brownian motion
obeys the diffusion equation. 
This correspondance extends
to higher dimensions for \CTPRW{s}
(CTPRW) as introduced 
by \citet*{masoliver1989}. 
%Masoliver, Lindenberg and Weiss \cite{masoliver1989}.
Persistent random walks in any dimensions can thus serve as
stochastic models from which the properties of radiative transfer
are obtained. 
In this work, I use these models as processes 
in \emph{phase space} and discuss the solution for the 
radiance $\phi(t,\,\vr,\,\uu)$ when the 
source emits a pulse at $t=0$. 

\subsection{Discrete-time persistent random walks}
% Taylor-Furth
The first mention of a persistent random walk dates back to the 
works by \citet{furth1920} and that of \citet{taylor1922},
who defined a one-dimensional persistant random walk as a
sequence of steps of a fixed length~$b$, occuring at a regular time pace.
%Note that this differs from the continuous persistence as
%designed by Ornstein \cite{ornstein1919}.
In modern language, the 
random process described by Fürth and Taylor is a Markov chain 
of states $(x_n,\,\dir_n)$ where $x_n$ is the 
position coordinate of the walker after $n$ steps and 
$\dir_n=\pm1$ is the direction of the next displacement~:
\begin{equation}
x_{n+1}=x_n+b_n\,\dir_n,
\label{xn}
\end{equation}
where $b_n>0$ is a sequence of step lengths (constant equal to $b$ in the
case of Fürth and Taylor's definition)
and $\dir_{n+1}=\,-\dir_n$ with probability $p$, or $\dir_{n+1}=\dir_n$
with probability $1-p$.
Fürth and Taylor's process is characterized by the transport mean free 
path~$\lp=\lim_{n\to\infty}\langle\dir_0x_n\rangle=b/2p$, 
(with $x_0=0$)
which is the average length
travelled without flipping direction.

Using this representation of the photon's
state, the collision (scattering event) of a photon with a scatterer is 
represented by a change
of its state from $(x,\,u)$ into $(x,\,u')$. 
The walk alternates between steps,
where the position is updated by Equation~\eqref{xn},
%while keeping~$u$ fixed
and such collisions.

% Masoliver-Weiss
\subsection{Continuous time persistent random walks}
Inspired by seminal remarks from 
\citet{kac1974} and the works of %DeWitt-Morrette and Foong
\citet{dewitt-morette1989} about the telegrapher's equation,
%a closely related problem~:\
%Masoliver, Lindenberg and Weiss 
\citet*{masoliver1989}
introduced the \CTPRW{} (CTPRW).
In one dimension, this random walk evolves according to the
Equation~\eqref{xn}, with randomly distributed and independent step 
lengths~$b_n\sim\Exp(\ell)$, which means that
each length is a random variable with
an exponential probability distribution function (PDF) $\Exp(\ell)$~:
$\mathrm{PDF}(b_n)=\frac1\ell\exp\big(-\frac{b_n}\ell\big)$.

In one dimension, the emission term is simply 
$\mathcal{I}[\phi]=(1-p)\phi(t,\,x,\,u)+p\phi(t,\,x,\,-u)$ and the \RTE{} is
\begin{equation*}
\left(\partial_t+cu\partial_x+p\frac c\ell\right)\phi(t,\,x,\,u)=
 p\frac c\ell\phi(t,\,x,\,-u)+S(t,\,x,\,u).
\end{equation*}
Remarking that, in this equation, $\ell$ only appears 
associated with $p$ in $\ell/p$, one deduces that
the statistics of the generalized coordinate $\Xg=(x,\,\dir)$ 
as a function of time $t$
only depend on the transport mean free path $\lp=\ell/2p$
and therefore that combinations of $\ell$ and $p$ 
yielding the same transport mean free path
$\lp$ are physically undistinguishable. 
In this work, I use $p=1$, which has the benefit of simplifying
the algebra without losing generality, such that $\ell$
is the only physical parameter of the process and 
the sequence $u_n$ is simply 
\begin{equation}
\dir_n=(-1)^n\dir_0.
\label{dn}
\end{equation}
Thanks to the Equation~\eqref{dn}, 
the analytic solution can easily be interpreted in
terms of parity of the number of scattering events
\citep{foong1992,foong1994}. 

%Even though the telegrapher's equation has been considered as a 
%candidate solution for "relativistic diffusion", it is only an approximate 
%solution of equations involving a diffusivity constant \cite{dunkel2009}.
In higher dimensions, the persistent random walk is a sequence of
steps of independent lengths $b_n\sim\Exp(\ell)$, and 
independent directions~$\uu_n$ uniformly distributed on the unit sphere.
In all dimensions, the Fokker-Planck equation of these processes 
\citep{rossetto2017} is Chandrasekhar's radiative transfer 
equation~\eqref{RTE}. %\cite{chandrasekhar}.
Therefore, a CTPRW process microscopically 
describes photons in a cloud of scatterers, 
observed in a frame~$\Ref$ where the cloud is at rest,
while the \RTE{} describes the phenomenon macroscopically.

\section{Solution in one dimension}
\label{sec:sol1}
Let me consider 
the propagation of light in one dimension after a source emits
a pulse of photons in the direction~$\dir_0$ at $t=0$
is
observed from a moving frame $\Ref'$, in uniform translation 
with respect to $\Ref$.
I denote by $v=\beta c$ the velocity of $\Ref$ observed in 
$\Ref'$.
Geometrical quantities in $\Ref'$
are labelled with a prime and the Lorentz factor is denoted by
$\gamma=1/\sqrt{1-\beta^2}$.

\subsection{Asymmetry induced by relative motion}
I consider a photon moving toward a scatterer located at a
distance $r'$ in $\Ref'$~:
the scatterer is reached by the photon after a time $r'/(c-\dir\,v)$.
In $\Ref$, the initial distance is $r=\gamma^{-1}r'$ and the travel 
time is $t=\gamma^{-1}r'/c$, such that, when $r'$ is the distance
between two scattering events, its average is the mean free path
in~$\Ref'$ and depends on the
direction of propagation of the photon. Writing 
$\ell'_\dir=\ell'_{\pm1}=\ell'_\pm$ gives
\begin{equation*}
 \frac{1}{\ell'_\pm}=\frac{c\mp v}{c}\frac{\gamma}{\ell}=
   (1\mp\beta)\frac\gamma{\ell},
\label{ell}
\end{equation*}
and consequently $\ell'_+\ell'_-=\ell^2$ is Lorentz invariant.

As observed from the moving frame~$\Ref'$, the photon's
random walk in the cloud of scatterers is an 
asymmetric persistent random walk, which is
a CTPRW with different transport mean free paths
depending on the direction of propagation~$\dir_n$~:
%\vspace*{-0.5cm}
\begin{equation}
b_n\sim\Exp(\ell'_{\dir_n}).
\label{an}
\end{equation}

\subsection{General solution}
The solution of the asymmetric persistent random walk has been recently
obtained without invoking special relativity~\citep{rossetto2018}. 
The solution follows entirely from the 
Equations~\eqref{xn}, \eqref{dn} and \eqref{an}, and is given in 
the cited Reference in terms of the asymmetry factors $\kappa$ and $\mu$:
\begin{equation*}
\kappa=\frac12\left(\frac1{\ell'_-}-\frac1{\ell'_+}\right)=
\frac{\beta\gamma}{\ell},\text{ and }
\mu=\frac{c}{2}\left(\frac{1}{\ell'_+}+\frac{1}{\ell'_-}\right)=
\frac{\gamma}{\ell}.
\end{equation*}
I here reproduce below the solution written in relativistic form:\\
%\centerline{(see Equation~\eqref{solution} \vpageref{solution})}
%\bigskip

\begin{strip}
\hrule
\begin{subequations}
\begin{eqnarray}
g(t';\,x',\,\dir\,|\,x'_0,\,\dir_0)=
\begin{cases}
\ds
\frac{ct'+\dir_0(x'-x'_0)}{\ell^2}\gev(x'-x'_0,\,t')
  +\dirac\big(ct'-\dir_0(x'-x'_0)\big)
  \exp\Big(-(1-\beta u_0)\gamma\frac{ct'}{\ell}\Big) & 
  \text{if }\dir=\dir_0,\\
\ds
\gamma\frac{1+\beta u_0}{\ell}\go(x'-x'_0,t') & \text{otherwise ;}
\end{cases}
\label{sol}\\
\text{with}\quad \go(x,t)=\frac 12\besselI_0(\xi)
    \exp\left(\frac{\beta\gamma x-\gamma c t}{\ell}\right)\Theta(ct-|x|) 
\quad\text{and}\quad
\gev(x,t)=\frac 1{2\xi} 
\besselI_1(\xi)
\exp\left(\frac{\beta\gamma x-\gamma c t}\ell\right)\Theta(ct-|x|),
\label{functions}
\end{eqnarray}
\label{solution}
\end{subequations}
\hrule
\end{strip}

%\begin{subequations}
%\begin{eqnarray}
%g(t';\,x',\,\dir_0\,|\,x'_0,\,\dir_0)= 
%\frac{ct'+\dir_0(x'-x'_0)}{\ell^2}\gev(x'-x'_0,\,t')
%\notag\\\quad
%  +\dirac\big(ct'-\dir_0(x'-x'_0)\big)
%  \exp\Big(-(1-\beta u_0)\gamma\frac{ct'}{\ell}\Big) \\
%g(t';\,x',\,-\dir_0\,|\,x'_0,\,\dir_0)= 
%\gamma\frac{1+\beta u_0}{\ell}\go(x'-x'_0,t') \quad
%\end{eqnarray}
%\begin{eqnarray}
%\text{with}\quad \go(x,t) =\frac 12\besselI_0(\xi)
%    \exp\left(\frac{\beta\gamma x-\gamma c t}{\ell}\right)\Theta(ct-|x|) \\
%\text{and}\quad
%\gev(x,t) =\frac 1{2\xi} 
%\besselI_1(\xi)
%\exp\left(\frac{\beta\gamma x-\gamma c t}\ell\right)\Theta(ct-|x|),
%\label{functions}
%\end{eqnarray}
%\label{solution}
%\end{subequations}

\noindent where $\xi=\sqrt{c^2t^2-x^2}/\ell$, 
$\besselI_n$ is Bessel's modified function of the first kind and 
$n$\textsuperscript{th} order,
and $\Theta$ is Heaviside's unit step function.
%As a function of $z$, $\gev_z$ and $\go_z$ are unnormalized moment generating
%functions of the number~$n$ of scattering events in the case where $n$ is
%even or odd, respectively.

The variable~$\xi$ is Lorentz invariant and the terms in the exponentials of 
Equation~\eqref{functions} corresponds to the change of coordinates between
$\Ref$ and $\Ref'$. It follows that global 
Lorentz covariance is satisfied by $g \dd x$ thanks to the invariance
of $\big(ct'+\dir_0(x'-x'_0)\big)\dd x'=\big(ct+\dir_0(x-x_0)\big)\dd x$.
This change of coordinates shows that the solutions of 
the symmetric persistent random walk in
$\Ref$ and the asymmetric persistent random walk in $\Ref'$ 
are related by a Lorentz transformation. 

%One of the reasons the telegrapher's equation has been dismissed as
%a candidate for relativistic Brownian motion is that its solution
%in space violates Chapman-Kolmogorov's relation.

\section{Chapman-Kolmogorov's relation}
\label{sec:cpr}
As $g$ is a probability density function, it is normalized according to
\begin{equation*}
\int_{\Xg'} g(t;\,\Xg'\,|\,\Xg'_0)=1 
\end{equation*}
where the symbolic integral 
$\int_{\Xg'}$ is a shortcut for $\int_{x'}\sum_{\dir=\pm1}$
expressing the integration on the whole phase space.
Note that the normalization depends on $\dir_0$. 

%The solutions discussed by Kac \cite{kac1974} and Dunkel and Hänggi
%\cite{dunkel2009} also have mixed initial conditions.% and is thus a
%superposition of two distinct solutions. 
Consider now a CTPRW
after a travel time $t_1$: The probability that $t_1$ coincides
with a scattering event is zero such that $u(t_1)$ is well defined.
Moreover the length of the step being
performed at time $t_1$ that remains to be travelled is 
distributed according to $\Exp(\ell_{u(t_1)})$ exactly as if this
was the first step of a new CTPRW starting at time $t_1$. 
This last property is called the
\emph{memorylessness} of the exponential distribution. These observations
imply that the generalized coordinate~$\Xg=(x,\,\dir)$ of the CTPRW follows a 
Markov process. 
Chapman-Kolmogorov's relation results from the Markovian
nature of the CTPRW process:
\begin{equation}
\int_{\Xg'}g(t_1;\,\Xg\,|\Xg')g(t_2;\,\Xg'\,|\Xg_0)=g(t_1+t_2;\,\Xg\,|\,\Xg_0).
\label{ChapmanKolmogorov}
\end{equation}
It follows immediately that if
the source is the distribution $S=\dirac(t)\phi_0(\Xg_0)$ 
then the radiance is the superposition 
\begin{equation*}
\phi(t,\,\Xg)=\int_{\Xg_0}g(t;\,\Xg\,|\,\Xg_0)\phi_0(\Xg_0),
\label{phi}
\end{equation*}
and so, the solution~\eqref{solution} has the properties of Green's functions.

In the relation~\eqref{ChapmanKolmogorov}, 
relativistic random walkers are characterized
not only by their position, but also their momentum. Classical random walks
are usually Markov point processes, in which the available transitions 
and their probabilities only depend on the current spatial position of
the walker. The relation~\eqref{ChapmanKolmogorov} suggests that 
relativistic random walks
should be represented as processes \emph{in phase space}.
If the source is not localized on a single point in phase space
then the process is a superposition of 
two independent Markov processes, it can therefore 
not be represented as a single point in phase space.
As an example, consider the
irradiance solution for an isotropic source
\begin{equation}
h(t;\,x\,|\,x_0)=\frac12\sum_{\substack{u_0=\pm1\\u=\pm1}}
  g(t;\,x,\,u\,|\,x_0,\,u_0),
\label{irr}
\end{equation}
which appears in References
\citep{goldstein1951,hemmer1961,kac1974,dunkel2009} and
\citep[p. 865]{morsefeschbach1}. 
The convolution 
%\begin{equation*}
$\int_{x'} h(t_1;\,x\,|\,x')h(t_2;\,x'\,|\,x_0)$
%\end{equation*}
contains superfluous terms such as 
\begin{equation*}
\int_{x'} g(t_1;\,x,\,+1\,|\,x',\,\underline{+1})
          g(t_2;\,x',\,\underline{-1}\,|\,x_0,\,+1),
\end{equation*}
that are non physical because the directions of propagation at position~$x'$  
in the two Green's functions 
(underlined in the above Equation)
do not match. This is the reason why~$h$ 
does not satisfy Chapman-Kolmogorov's relation.

\section{First-passage time statistics}
\begin{figure}[tb]
\begin{center}
\includegraphics[height=3.8cm]{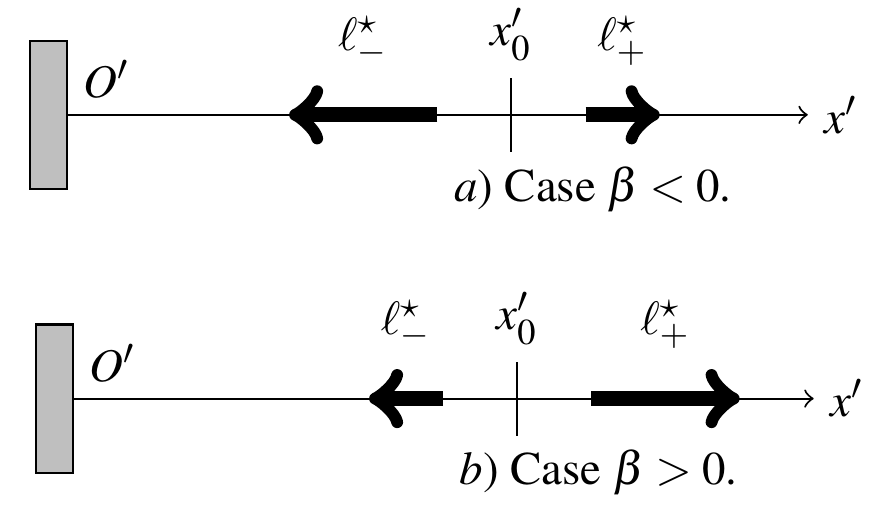}
\caption{\label{fig:APRW}The two cases of the asymmetric persistent 
random walk in one dimension.
{\it a}) The case $\beta<0$. 
In this case, $\ell_->\ell_+$, the probability of passage at the origin 
$O'$ is equal to~$1$ and the average first-passage time is finite.
{\it b}) The case $\beta>0$.
In this case $\ell_+>\ell_-$, and the photon has a finite
probability of never visiting the origin~$O'$.
}
\end{center}
%\vspace*{-0.5cm}
\end{figure}

Numerous applications of random walks in Physics involve
properties like local time, integrated area and 
first-passage times \citep{majumdar2005,rossetto2013}.
In the case of radiative transfer, the first-passage time distribution
is the signal measured after the emission of a pulse localized on a single 
point in phase space.
A~device recording such a pulse indeed measures a signal
proportional to the first-passage time distribution at its position.
Let me then consider the first-passage problem at the 
origin~$O'$ in $\Ref'$ when the CTPRW starts at~$x'_0>0$. 
%The origin $O'$ of $\Ref'$ is moving in $\Ref$ with velocity~$v$.
As a consequence of the 
asymmetry of mean free path, 
a photon observed from $\Ref'$ experiences an average drift.
The Figure~\ref{fig:APRW} displays
an illustration of the two possible cases.
In the first case ($\beta<0$, Figure~\ref{fig:APRW}{\it a}), 
%the observer is moving toward the source and
the photon is drifted toward $O'$~; 
the distribution of the first-passage time at the
origin~$P_1(t'\,|\,\Xg'_0)$ has finite moments,
as opposed to the case where $\beta\geq0$.
In the second case ($\beta<0$, Figure~\ref{fig:APRW}{\it b}),
%the observer is moving away from the source and 
the photon is drifted away from $O'$~; 
the probability $R(\Xg'_0)=\int_0^\infty P_1(t'\,|\,\Xg'_0)\dd t'$ 
that the photon starting at position $x'_0$ 
ever reaches~$O'$ departs
from~1 as opposed to the case where $\beta\leq0$. 
The expressions of $P_1$ and $R$ are given in the Table~\ref{FP} below.

\begin{strip}
\begin{center}
\vskip-0.7cm
\hrule
\vskip0.2cm

\captionsetup{type=table}
\caption{\label{FP}Table of main results for the
first-passage time statistics. $P_1(t'\,|\,\Xg'_0)$ is
the distribution of the first passage time at the origin $O'$ and
$R(\Xg'_0)$ is the probability that the photon will reach the origin
in a finite time. 
$R(\Xg'_0)$ is equal to~$1$ whenever $\beta\leq0$. }
\def\arraystretch{1.2}%
\begin{tabular}{|l|l|l|}
\hline &
\multicolumn{1}{c|}{all values of $\beta$} & 
\multicolumn{1}{c|}{$\beta>0$} 
\\
\hline
$\dir_0=-1$ & 
$\ds P_1(t'\,|\,x'_0,\,-1) = 2x'_0\gev(-x'_0,\,t')
   +\dirac(ct'-x'_0)\exp(-(1+\beta)\gamma x'_0/\ell)$
&
$\ds R(x'_0,\,-1) = \exp(-2\beta\gamma x'_0/\ell)$ %&
\\
$\dir_0=+1$ & 
$\ds P_1(t'\,|\,x'_0,\,+1) = \frac{2(1-\beta)\gamma}{\ell(ct'+x'_0)}
   \big(x'_0\go(-x'_0,\,t')+(ct'-x'_0)\gev(-x'_0,\,t')\big)$ &
$\ds R(x'_0,\,+1)= \frac{1-\beta}{1+\beta}\exp(-2\beta\gamma x'_0/\ell)$ %& 
\\
\hline
\end{tabular}
\end{center}
\end{strip}

\subsection{An effective "flip" process}
%The initial direction of propagation~$\dir_0$ of the photon
%plays an important role and changes the first-passage time
%statistics in both cases $\beta>0$ and $\beta<0$. 
As shown in the Reference \citep{rossetto2018},
the Laplace transforms~$\TL P_1$ of the first-passage time distributions
for~$\dir_0=\pm1$ are related by
\begin{gather}
\TL P_1(s\,|\,x'_0,\,+1) = \TL P_1(s\,|\,x'_0,\,-1) \TL f(s), 
\label{flip}\\
%\begin{equation}
\text{where }
  \TL f(s) =
   \frac{(1-\beta)\gamma}{\frac\ell cs+\gamma
     +\sqrt{(\frac\ell c s+\gamma)^2-1}}. \quad
\label{F}
\end{gather}
The equation~\eqref{flip} shows that the first-passage time process of
a CTPRW with initial direction $\dir_0=+1$
is the convolution of the first-passage time process of a CTPRW with 
$\dir_0=-1$ 
with a new process $F$ that I call a "flip process".

A CTPRW with initial direction $\dir_0=+1$ 
has the following effective
behavior: it first spends a certain time to "flip" (that is: to set itself
in the same initial conditions as a CTPRW process with $\dir_0=-1$)
then it follows the course of a regular CTPRW 
with $\dir_0=-1$ until it reaches the origin.
As $\TL f$ is independent of $x'_0$, the "flip" is completely independent
of the initial distance to the origin and its effect cannot be
interpreted as, or reproduced by, a shift of the starting position $x'_0$. 

The photons emitted away from $O'$
(with $\dir_0=+1$) 
are thus observed by the moving operator in $O'$
with a random supplementary delay corresponding to the "flip time".

\subsection{Statistics of the "flip" process}
The inverse Laplace transform of~$\TL f$ 
is the distribution of time until the "flip" occurs.
Using Reference \citep[formul\ae{} 29.3.52 and 29.3.53]{abramowitzstegun}, one
obtains from Equation~\eqref{F}
%\vspace*{-0.3cm}
\begin{equation}
f(t)=(1-\beta)\gamma \,\frac{\besselI_1(ct/\ell)}{t}
  \e^{-\gamma ct/\ell}.
%=P_1(t\,|\,0,+1)
\end{equation}
Up to my knowledge, this expression is not a referenced 
probability distribution.
Note that $f(t)=P_1(t\,|\,0,+1)$ so that, when $x'_0=0$, the "flip"
time is also the time of return at $O'$. 
For $\beta<0$, the mean "flip" time and its variance are
\begin{equation}
\Esp[F]=\frac{1}{|\beta|}\frac{\ell}{\gamma c},\qquad
\Var(F)=\frac1{|\beta|^3} \left(\frac{\ell}{\gamma c}\right)^2.
\end{equation}
More properties of~$F$ are given in the Appendix~\ref{anx:F}.

\begin{figure}
\begin{center}
\includegraphics[width=0.9\linewidth]{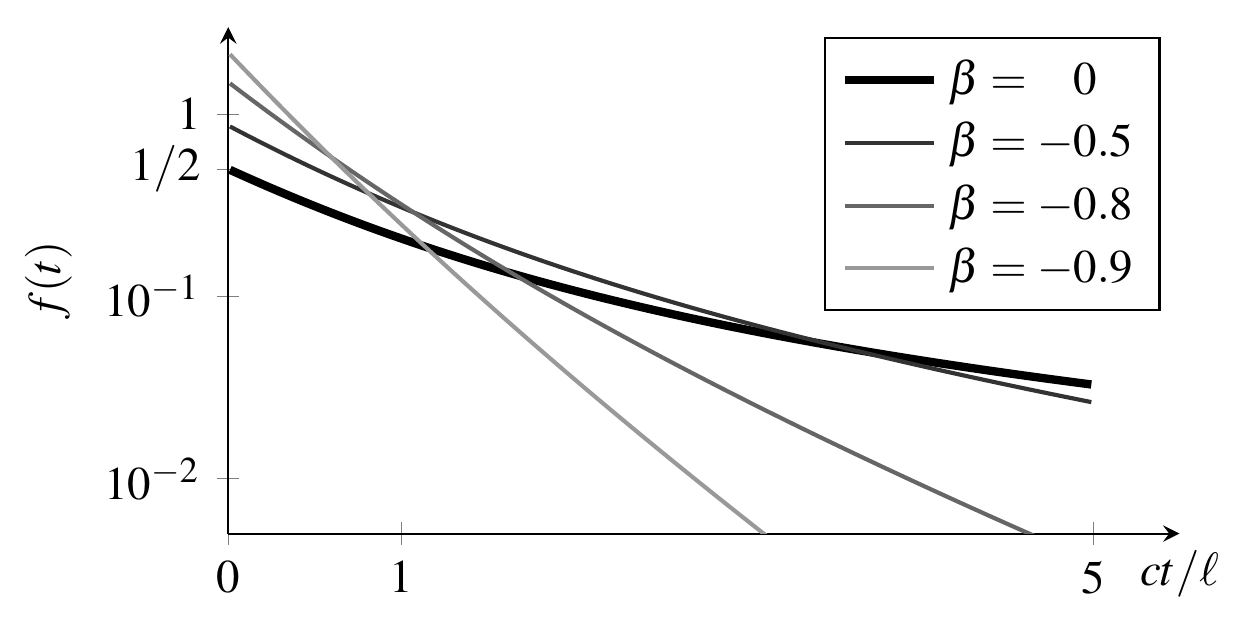}
\caption{\label{fig:switch}Graphs of the distribution of the 
"flip" process for some values of $\beta\leq0$. For $\beta=0$,
the distribution is normalized, but it has no finite moments.
As $\beta$ approaches $-1$, the time distribution gets more concentrated
near $t=0$.
}
\end{center}
%\vspace*{-0.5cm}
\end{figure}

\section{Lorentz covariance in two dimensions}
\label{sec:lc2}
As previously suggested in the case of the one-dimensional persistent random
walk, relativistic random walks should be constructed
as Markov processes in phase space. In two dimensions,
a point in the photon's phase space is a pair $(\vr,\,\uu)$ where 
$\vr$ is the two-dimensional position and $\uu$ is the two-dimensional
unit vector of the direction of propagation.
The exact solution of the 
two-dimensional CTPRW 
in the reference frame where the cloud of scatterers is at rest, 
$\Ref$, 
with initial conditions
$\Xg_0=(\vr_0=0,\,\uu_0)$ at $t=0$ was recently published
\citep{rossetto2017}. It essentially depends on the Lorentz
invariant $c^2t^2-\vr^2$ and on the variable $\XL$: 
%\vspace*{-0.1cm}
\begin{equation}
\XL^2=\frac{2}{\ell^2}
   \frac{(ct-\vr\cdot\uu)(ct-\vr\cdot\uu_0)}{1-\uu\cdot\uu_0}
   -\xi^2
\label{X}
\end{equation}
($\xi=\sqrt{c^2t^2-\vr^2}/\ell$) 
and it decomposes as the sum $g=g_0+g_1+g_\infty$ where
$g_0$ is the unscattered contribution (Dirac delta function),
$g_1$ is the single scattering contribution and 
%\begin{widetext}\[
\begin{multline*} 
g_\infty(t,\,\Xg\,|\,\Xg_0)=
   \frac{1}{2\pi{\ell}^2}
   \frac{\e^{-ct/\ell}}{1-\uu\cdot\uu_0}
 \Theta(ct-\|\vr\|) \times\dots \\
  \cdots \Re\left[\Eint(\ii\XL)\e^{\ii\XL} 
           -\Eint\left(\ii\XL-\xi\right)
            \e^{\ii\XL}\right] % \Theta(ct-\|\vr\|),
\label{tr2d}
%\]\end{widetext}
\end{multline*}
is the multiple scattering contribution, 
with $\Eint$ the exponential integral function
\citep[Chapter 5]{abramowitzstegun}.

As the process is Markovian, 
Chapman-Kolmogorov's relation is 
fulfilled and the expression~\eqref{ChapmanKolmogorov} is valid  
using the convention that, in two dimensions, the phase space integral 
$\int_{\Xg}$ denotes 
$\int_{\vr}\int_{\uu}$.
Again, Chapman-Kolmogorov's relation is satisfied for all
pulse initial conditions \emph{in phase space}
separately, but not necessarily for their superpositions.

The velocity of $\Ref$ in $\Ref'$ is denoted, 
without loss of generality, by $\vv=v\ue_x$. 
The coordinates of $\vr$ transform into the coordinates
of $\vr'$ by a Lorentz boost
whereas the coordinates of $\uu$ and $\uu_0$ transform 
according to the addition law of velocities.
Using these relations and elementary algebra operations
shows that the variable~$\XL$ 
is Lorentz invariant
(see the Appendix for details).
The Lorentz covariance of the term $g_\infty$ follows
and that of the terms~$g_0$ and $g_1$
is straightforward. 
%The two-dimensional solution~\eqref{tr2d} therefore is the 

%Note that the normalization of~$g_{\infty}$
%is obtained with the area element transformation 
%$\dd\vr=\gamma^2(1-\beta u_x)(1-\beta u_{0x})\dd\vr'$.

\section{The Brownian limit}
As mentioned in the introduction, the telegrapher's equation also 
appears as a modification of Smoluchowski's diffusion 
equation \citep{brinkman1956,sack1956}, which microscopic model
is Brownian motion.
%The CTPRW process is thus an extension of Brownian motion.
The classical constructions  
of Brownian motion are, in various ways, mathematical limits
of an underlying random walk performing small steps
at a rate going to infinity (see for instance the works
of \citet{wiener1923} and \citet{ito1950}).
This limit implies that 
the instantaneous speed of a particule
performing the random walk is infinite.
Although the limit of infinite speed is legitimate 
in classical physics, it does not comply with special relativity. 
Arguably, a relativistic counterpart of the classical Brownian 
motion should involve particles moving at largest possible
speed, the speed of light $c$. Such particles then must be massless.
The relativistic Brownian motion of massive particles 
has been extensively studied by Dunkel
and Hänggi \citep{dunkel2005a,dunkel2005b,dunkel2009} using stochastic
differential equations, where the limit of infinite rate 
pertains to the concept of noise.

The Brownian limit of the
general solution~\eqref{solution} corresponds to taking the speed
of light $c$ to infinity and the mean free path $\ell$ to zero, while
keeping the product $\ell c=2D$ constant. It yields
a Gaussian distribution in $\Ref'$ centered at $\beta ct'=vt'$
and of variance $2Dt'$. 
The contributions of the initial direction of
propagation $\dir_0$ vanish in the Brownian limit of 
Equation~\eqref{solution}.
The two cases~$\dir_0=\pm1$ in the Equation~\eqref{sol} are therefore 
undistinguishable 
such that the solution $g(x',\,t')$ in $\Ref'$ is 
their sum. 
Lorentz invariance (and thus causality) also
disappears in this limit.

In the limit $c\to\infty$, the Lorentz factor 
is $\gamma\simeq1+\frac12\beta^2$. 
The expansion~$\xi\simeq \frac{ct}\ell-x^2/2ct\ell$
and the asymptotic form $\besselI_\nu(x)\simeq\e^x/\sqrt{2\pi x}$
\citep[formula 9.7.1]{abramowitzstegun}
give
\begin{equation*}
%\begin{split}
g(x',\,t') \simeq
  \frac{\exp\big(\frac{x'^2}{4 c^2t'^2}\big)}{\sqrt{2\pi ct'\ell}}
  \exp\left[-\frac{(x'-\beta ct')^2}{2\ell ct'}\right]
  \Theta(ct'-|x'|).
%\frac1\ell\go_1(x',\,t')&\simeq
  %\frac{1}{2\ell} 
  %\frac{\exp\big(\frac{x'^2}{4 c^2t'^2}\big)}{\sqrt{2\pi ct'/\ell}}
  %\exp\left[\frac{ct'}\ell-\frac{x'^2}{2\ell ct'}+
  %\beta\frac{x'}\ell-\Big(1+\frac{\beta^2}2\Big)\frac{ct'}\ell\right]
  %\Theta(ct'-|x'|)\\
  %&\simeq 
  %\frac12 \frac{1}{\sqrt{2\pi\ell ct'}}
  %\exp\left[\frac{x'^2}{4c^2t'^2}-\frac{1}{2\ell ct'}
    %\left(x'-\beta ct'\right)^2\right]
  %\Theta(ct'-|x'|);\\
%\frac1\ell\gev_1(x',\,t')  &\simeq 
  %\frac12 \frac{1}{\sqrt{2\pi\ell ct'}}
  %\exp\left[\frac{3x'^2}{4c^2t'^2}-\frac{1}{2\ell ct'}
    %\left(x'-\beta ct'\right)^2\right]
  %\Theta(ct'-|x'|).\\
%\end{split}
\end{equation*}
The total solution is, in the limit $c\to\infty$ and $\ell\to0$
with $c\ell=2D$ constant, equal to 
\begin{equation*} 
g(x',\,t')=\frac{1}{\sqrt{4\pi Dt'}}\exp\left[-\frac{(x'-vt')^2}{4Dt'}\right].
\end{equation*}

\section{Closing remarks}
In this Letter, I have shown that the solutions of the radiative
transfer equation are naturally Lorentz covariant and can be
obtained from the solutions expressed in phase space.
I proved that these solutions satisfy the transitivity 
of Green's functions only in phase space,
not in position space.
In one dimension, 
I have unveiled a "flip process" that translates as a delay
in the measured signal for photons emitted away from the observer.
I also have shown that the moments of a signal measured from a pulse 
become finite in the case where the observer is moving toward the source
($\beta<0$).
These results have been obtained thanks to the stochastic model 
of continous-time persistent random walks (CTPRW), which are Markov processes 
in phase space. 
I showed that the Markov property of the CTPRW follows from the 
memorylessness
of the exponential probability distribution
and the phase space representation.

One should naturally expect that 
the same approach applies in three dimensions.
The three-dimensional Green's function would indeed 
be of interest in astrophysics
and medical imaging, but no exact solutions are known in three dimensions,
even for the energy density (the process in position space).
It is however possible that a solution for the radiance, {\it i. e.} in 
phase space, exists.
Such a solution could be expressed in terms of a 
Lorentz invariant variable such as $\XL$ from Equation~\eqref{X},
which three-dimensional
counterpart also is Lorentz invariant.
In any event, I believe that this 
Letter will trigger progress in this direction.

Using the limit $c\to\infty$, I showed that 
the CTPRW is an extension of Brownian motion 
and therefore that radiative transfer
is a natural extension of diffusion in special relativity, requiring a 
solution in phase space.
Several works have already suggested to consider 
relativistic Brownian motion
as a process in phase space \citep{dudley1966,hakim1968}.
A straightforward extension of the CTPRW for massive particles would,
for instance, be
a process selecting a random momentum
according the Jüttner-Maxwell distribution \citep{juttner1911}, 
and a random, exponentially distributed, free travel distance at each step.
Such a construction could avoid the difficulties related to the
interpretation of stochastic integrals in the abovementioned works.

%\appendix

\section*{Appendices}
\renewcommand\thesubsection{\Alph{subsection}}
\subsection{Lorentz invariance of $\XL$}
\label{AppX}
Here is a proof that the variable $\XL$ defined
by the Equation~\eqref{X} is a relativistic invariant.
In two dimensions, the space-time coordinates of $\vr=(x,\,y)$ transform as
\begin{equation*}
x  = \gamma x' - \beta\gamma ct', \qquad
y  = y', \qquad
ct = \gamma ct' - \beta\gamma x' , 
\end{equation*}
whereas the components of the direction of propagation $\uu=(u_x,\,u_y)$
transform through the velocity addition law~:
\begin{equation*}
 u_x  = \frac{u'_x-\beta}{1-\beta u'_x}, \qquad
 u_y  = \frac{u'_y}{\gamma(1-\beta u'_x)}.
\end{equation*}
Therefore we obtain
\begin{equation*}\begin{split}
   ct-\vr\cdot\uu &= \gamma ct'-\beta\gamma x'
    -\frac{(\gamma x' -\beta\gamma ct')(u'_x-\beta)
    +y' u'_y/\gamma}{1-\beta u'_x}\\
%  &=\frac{\gamma ct'
%          {\color{rouge}-\beta\gamma u'_xct'}
%          {\color{bleu}-\beta\gamma x'}
%          +\beta^2\gamma u'_xx'
%          -\gamma x'u'_x
%          {\color{bleu}+\beta\gamma x'}
%          {\color{rouge}+\beta\gamma ct'u'_x}
%          -\beta^2\gamma c't
%          - y'u'_y/\gamma}{1-\beta u'_x},\\
%  &=\frac{(1-\beta^2)\gamma^2 ct'
%          -(1-\beta^2)\gamma^2 x'u'_x
%          - y'u'_y}{\gamma(1-\beta u'_x)}
  & = \frac{ct'-\vr'\cdot\uu'}{\gamma(1-\beta u'_x)}.
\end{split} 
\end{equation*}
%Consequently, the numerator transforms as
%\begin{multline*} (ct-\vr\cdot\uu)(ct-\vr\cdot\uu_0)=\\
%   \frac{1-\beta^2}{(1-\beta u'_x)(1-\beta u'_{0x})}
%(ct'-\vr'\cdot\uu')(ct'-\vr'\cdot \uu'_0).
%\end{multline*}
Moreover, 
\begin{equation*}
 \uu\cdot\uu_0  =
 \frac{u'_x-\beta}{1-\beta u'_x}\frac{u'_{0x}-\beta}{1-\beta u'_{0x}}
 +\frac{u'_y}{\gamma(1-\beta u'_x)}\frac{u'_{0y}}{\gamma(1-\beta u'_{0x})},
\end{equation*}
and so
\begin{equation*}
\begin{split}
1-\uu\cdot\uu_0
% &=\frac{(1-\beta u'_x)(1-\beta u'_{0x})
%        -u'_xu'_{0x}+\beta u'_x +\beta u'_{0x}-\beta^2
%        -u'_yu'_{0y}/\gamma^2}{(1-\beta u'_x)(1-\beta u'_{0x})}, \\
% &=\frac{1
%        {\color{rouge}-\beta u'_x}
%        {\color{bleu}-\beta u'_{0x}}
%        +\beta^2 u'_xu'_{0x}
%        -u'_xu'_{0x}
%        {\color{rouge}+\beta u'_x}
%        {\color{bleu}+\beta u'_{0x}}
%        -\beta^2
%        -u'_yu'_{0y}/\gamma^2}
%        {(1-\beta u'_x)(1-\beta u'_{0x})}, \\
% &=\frac{1-\beta^2-(1-\beta^2) u'_xu'_{0x}-u'_yu'_{0y}/\gamma^2}
%        {(1-\beta u'_x)(1-\beta u'_{0x})}\\& 
=
   \frac{1-\beta^2}{(1-\beta u'_x)(1-\beta u'_{0x})}(1-\uu'\cdot\uu'_0).
\end{split}
\end{equation*}
As a conclusion, the fraction in Eq.~\eqref{X} transforms as
\begin{equation*}
 \frac{(ct-\vr\cdot\uu)(ct-\vr\cdot\uu_0)}{1-\uu\cdot\uu_0}
 = \frac{(ct'-\vr'\cdot\uu')(ct'-\vr'\cdot\uu'_0)}{1-\uu'\cdot\uu'_0}.
\end{equation*}
This proves the announced invariance of $\XL$.

\subsection{More properties of $F$}
\label{anx:F}
The probability density of the "flip" time is, to my knowledge,
not a referenced probability density function. I give here 
the expressions of the probability density~$\sigma_n$ of the sum 
of $n\geq1$ independent
"flip" processes $F_1+F_2+\cdots + F_n = \Sigma_n$ and the 
moments of order~$k$ of $\Sigma_n$. These are established using the
relation $\sigma_n=f^{\otimes n}$ and its Laplace transform
$\TL \sigma_n=\big(\TL f\big)^n$, where $\otimes$ denotes 
convolution and
${}_2F_1$ is Gauss's hypergeometric function.
\begin{subequations}
\begin{eqnarray}
\sigma_n(t)&=& %f^{\otimes n}(t)=
  n\big[(1-\beta)\gamma\big]^n \;\frac{\besselI_n(ct/\ell)}{t}
  \e^{-\gamma c t /\ell},\\
\Esp\big[\Sigma_n^k\big]&=&
\left(\frac{1-\beta}{2\gamma}\right)^n\,
\left(\frac{\ell}{\gamma c}\right)^k\,
\frac{(n+k-1)!}{(n-1)!} \dots \notag\\
&&\qquad\dots\;
{}_2F_{1}\!\left(\begin{array}{c}\frac{n+k}{2}\;
                               \frac{n+k+1}{2}\\n+1\end{array}\;
\vrule\;\frac{1}{\gamma^2}\right).
%)big\langle t^k\big\rangle_{S^{\otimes n}}&=&(1-\beta)\frac{k!}{2}\;
%\left(\frac{\ell}{\gamma c}\right)^k
%{}_2F_{1}\!\left(\begin{array}{c}\frac{k+1}{2}\;\frac{k+2}{2}\\2\end{array}\;
%\vrule\;\frac{1}{\gamma^2}\right),\\
\end{eqnarray}
\end{subequations}
\vskip3cm

%\begin{strip}
\bibliography{flip}
%\end{strip}
\end{document}